\documentclass[intlimits,twoside,a4paper]{article}
\usepackage{graphicx}
\usepackage{subcaption} 
\usepackage{float}
\usepackage{appendix}
\usepackage{listings}
\usepackage{xcolor}
\usepackage[cp1251]{inputenc}
\usepackage{enumitem}

\usepackage[eqsecnum]{cmpj3}
\issue{2025}{28}{3}{33601}
\doinumber{10.5488/CMP.28.33601}

\title[Algebraic solution and thermodynamic properties of graphene]%
{Algebraic solution and thermodynamic properties of graphene in the presence of minimal length}%

\author[J. Gb\`etoho, F. A. Dossa,  G. Y. H. Avossevou]
{J. Gb\`etoho\orcid{0009-0005-1145-0853}\refaddr{label1}, F. A. Dossa
 \orcid{0000-0002-2694-4144}\refaddr{label1}\thanks{Corresponding author: \email{dossafanselme@gmail.com}.},
 G. Y. H. Avossevou\orcid{0000-0002-9609-0340}\refaddr{label2}}
\addresses{
\addr{label1} Laboratory of Physics and Applications (LPA), Universit\'e Nationale des Sciences, Technologies,
Ing\'enierie et Math\'ematiques (UNSTIM) Abomey, BP: 2282 Goho Abomey, R\'epublique du B\'enin
\addr{label2} Institut de Math\'ematiques et de Sciences  Physiques (IMSP),
 Universit\'e d'Abomey-Calavi (UAC),
01 BP 613 Porto-Novo, R\'epublique du B\'enin}


\date{Received January 03, 2025, in final form March 05, 2025}

\begin{document}
\maketitle
\begin{abstract}
Graphene is a zero-gap semiconductor, where the electrons propagating inside are described by the ultra-relativistic Dirac equation normally reserved for very high energy massless particles. In this work, we show that graphene under a magnetic field in the presence of a minimal length has a hidden $su(1,1)$ symmetry. This symmetry allows us to construct the spectrum algebraically. In fact, a generalized uncertainty relation, leading to a non-zero minimum uncertainty on the position, would be closer to physical reality and allow us to control or create bound states in graphene. Using the partition function based on the Epstein zeta function, the thermodynamic properties are well determined. We find that the Dulong--Petit law is verified and the heat capacity is independent of the deformation parameter.

\printkeywords
%
% \pacs 03.65.Ge; 05.70.Ce; 03.65.Ge; 03.65.Ge (optional)
\end{abstract}

\section{Introduction}

Graphene has attracted the attention of scientific community in recent years due to its special pro\-per\-ties. This is the first two-dimensional crystalline solid that has been experimentally developed. This solid consists of a two-dimensional (2D) sheet composed of carbon atoms arranged in a hexagonal honeycomb lattice.
Its first theoretical study was carried out by Wallace~\cite{aa1}. In 2004, Novoselov {et al.}~\cite{aa2} succeeded in isolating and electrically contacting a graphene sheet from the mechanical exfoliation of natural graphite. A graphene sheet consists of carbon atoms arranged in a two-dimensional hexagonal network. It has very particular electronic and mechanical properties, which have attracted the interest of both the scientific community and industrialists. Indeed, graphene has a very high electronic mobility and a high stability at the nanometric scale. In addition, unlike nanotubes, graphene can be produced on large surfaces. All these properties make it possible to envisage the manufacture of graphene-based electronic systems that could process data $10$ times faster than the current systems. However, graphene does not have an energy gap, which is an essential condition for the manufacture of transistors. This discovery of graphene paved the way for the study of new 2D materials. Like graphene, many other 2D materials have been prepared experimentally. This is particularly the case for hexagonal boron nitride, which is a wide bandgap material, whose structural properties are similar to those of graphene.
It is noted that the study of graphene has attracted much interest in controlling or confining electrons in graphene due to its spectacular properties and potentially important applications in nanotechnology. Confinement attempts, using position-dependent mass \cite{aa12} and magnetic fields, have been made to create bound states in graphene. Recently, a series of studies concerning the interaction of graphene electrons have been conducted, in order to find a way to confine charges \cite{aa17,aa21,aa22,Ba,Ba1,Ba2}. More recently, exact analytical solutions of the Dirac--Weyl equation in graphene under various magnetic fields in the Cartesian coordinate system have been found~\cite{cac}. Furthermore, in \cite{aa1a} the massless Dirac equation in the presence of a constant external magnetic field has been solved within the framework of the Dunkl formalism, where the Dunkl parameters modify the conventional results of graphene thermal quantities. Parity effects appear on the thermal quantities at low temperatures.

In this work, we study the Dirac--Weyl equation in graphene under magnetic fields with the ge\-ne\-ra\-li\-zed uncertainty principle. Indeed, the pointlike character of particles is a basic postulate in quantum mechanics; one of the fundamental consequences, which follows from it, is the localizability of particles. In fact, a generalized uncertainty relation, leading to a minimal non-zero uncertainty on the position, would be closer to physical reality. This could help to control or create bound states in graphene.
Several works have been done in recent years in the framework of the generalized uncertainty principle to show the impact of this principle on quantum mechanical problems~\cite{dos1,dos2}. We show that the Dirac--Weyl equation for the deformed graphene has a hidden $su(1, 1)$ symmetry. Thus, the eigenvalues and eigenstates are constructed algebraically. Next, we use the properties of the zeta functions to determine, from the partition function, thermodynamic quantities such as: free energy, entropy, specific heat and average energy.

\section{Basics of graphene}

\begin{figure}[htbp]
\begin{center}
  (a)\includegraphics[scale=0.6]{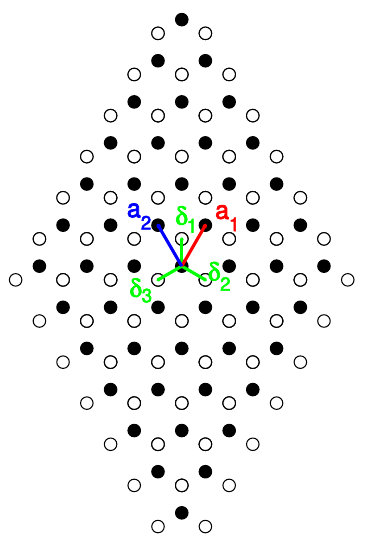}
   (b)\includegraphics[scale=0.55]{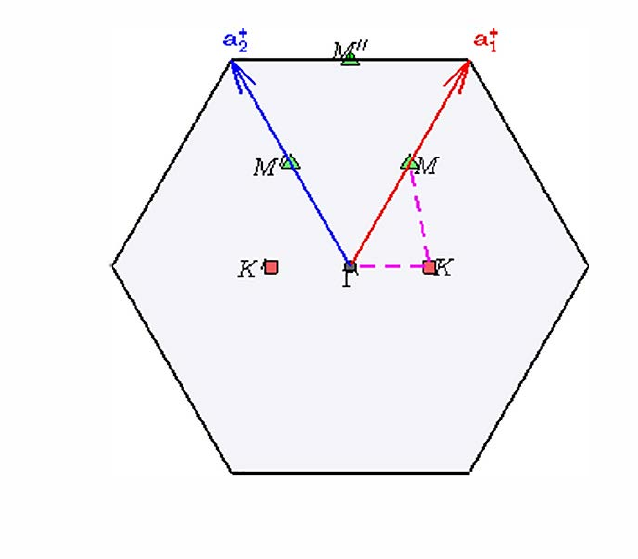}
 \end{center}
   \caption{(Colour online) (a) Graphene honeycomb lattice. ($\bullet:$ $A$ sublattice) and  ($\circ:$ $B$ sublattice).
   	The vectors $\delta_1$, $\delta_2$, and $\delta_3$ connect neighboring carbon atoms, separated by a distance $a = 0.142$ nm.
   	The vectors $\mathbf{a}_1$ and $\mathbf{a}_2$ form the basis of the triangular Bravais lattice.
   	(b) First Brillouin zone with band path. Its primitive vectors are $\mathbf{a}_1^\star$ and $\mathbf{a}_2^\star$.
   	The shaded region represents the first Brillouin zone, with its center~$\Gamma$, the inequivalent corners $K$ (red squares) and $K{'}$ (red squares), as well as the three crystallographic points $M$, $M{'}$, $M{''}$ (green triangles). The path $\Gamma\rightarrow K\rightarrow M \rightarrow \Gamma$ (dashed magenta line) is a standard line in solid-state physics to represent the electronic bands of graphene. Original representation, inspired by~\cite{mob}.}
  \label{kl}
  \end{figure}

Structurally, graphene consists of an ordered hexagonal carbon lattice. Although it has some symmetry, the resulting lattice is not a Bravais lattice, but a superposition of two triangular Bravais sublattices formed by atoms of types $A$ and $B$, as shown in figure~\ref{kl}(a). In the reciprocal lattice, the Brillouin zone figure~\ref{kl}(b) is the primitive cell of the wave vector space $\vec{k}$ and has the particularity of being capable of representing all the properties of the real lattice.

Inside this zone, we identify the points of high symmetry, also called critical points. The points $K$ and~$K'$ are of great importance in the description of the band structure. Graphene  has a band gap that is zero~\cite{aa20a}, so it can be assimilated to a semiconductor without a band gap. Its basic electronic properties are easily understood within a simple tight-binding model in which electron hopping is restricted to the nearest sites. The Hamiltonian resulting from a decomposition into Bloch states with a lattice momentum~$\vec k$ reads
\begin{eqnarray}\label{h}
	H=\xi\left(\begin{array}{cc}
		0& \gamma^\star_{\bf k}\\
		\gamma_{\bf k} & 0
	\end{array}\right)=v\left(\begin{array}{cc}
		0& p_x-\ri p_y\\
		p_x+p_y & 0
	\end{array}\right)=v\bf{\sigma}\cdot{\bf p},
\end{eqnarray}
which is obtained within a tight-binding model, where one considers the electronic hopping between nearest-neighbouring sites with a hopping amplitude $\xi$.
Here, the Fermi velocity $v$ plays the role of the velocity of light $c$, which is nevertheless roughly $300$ times larger, $c=300v$. The Pauli matrices ${\bf \sigma}= (\sigma_x, \sigma_y, \sigma_z)$ are given by
\begin{eqnarray}
 \sigma_x=\left(\begin{array}{cc}
		0& 1\\
		1 & 0
	\end{array}\right), \quad \sigma_y=\left(\begin{array}{cc}
		0& -\ri\\
		\ri & 0
	\end{array}\right), \quad \sigma_z=\left(\begin{array}{cc}
		1& 0\\
		0 & -1
	\end{array}\right).
\end{eqnarray}
The Hamiltonian (\ref{h}) is indeed formally the Hamiltonian of
massless 2D particles, and it is sometimes called Weyl or Dirac Hamiltonian. In order to describe free electrons in a magnetic field, one needs to replace the momentum by its gauge-invariant form
\begin{equation}
 {\bf p}\rightarrow {\bf p}-\frac{e}{c}{\bf A},
\end{equation}
where {\bf A} is the vector potential that generates the magnetic field
${\bf B}={\bf \nabla} \times {\bf A}$. This gauge-invariant momentum is proportional to the electron velocity
$v$, which must naturally be gauge-invariant because it is a physical quantity.

The Hamiltonian (\ref{h}) becomes
\begin{equation}\label{hb}
 H=v {\sigma}\cdot\left({\bf p}-\frac{e}{c}{\bf A}\right),
\end{equation}
where ${\bf A}$ is the electomagnetic vector potential and in the symmetric gauge is given by
\begin{equation}
 {\bf A}=\frac{B}{2}(-y, x, 0).
\end{equation}
We consider the Dirac electron moving in graphene under a static orbital magnetic field acting perpendicular to the graphene plane. Moreover, we disregard the electronic spin degree of freedom and the presence of any disorder in the system.
In order to analyse the Hamiltonian (\ref{hb}) in a quantum mechanical treatment, we use the standard method, the canonical quantization, where one interprets the physical quantities as operators that act on state vectors in a Hilbert space. These operators in general do not commute with each other, i.e., the order matters in which they act on the state vector that describes the physical system.
Under these circumstances, the low energy excitation of the electron around a Dirac point in the Brillouin zone is described by the following Dirac--Weyl equation
\begin{equation}\label{hgg}
 H\Psi(x,y)=v{\bf\sigma}\cdot\left({\bf p}-\frac{e}{c}{\bf A}\right)=E\Psi(x,y).
\end{equation}
The two-component wavefunction is the column matrix
$\Psi(x, y) = (\varphi(x, y), \varphi (x, y))^T$, with the super-index $T$ denoting the matrix transposition. From equation (\ref{hgg}) we can
write
\begin{eqnarray}\label{eqq}
 v\!\left(\begin{array}{cc}
		0& 2p_z+\ri\omega \bar z\\
		 2p_{\bar z}-\ri\omega z & 0
	\end{array}\right)\left(\begin{array}{c}
		\varphi_1\\
		\varphi_2
		\end{array}\right)=E\left(\begin{array}{c}
		\varphi_1\\
		\varphi_2
		\end{array}\right),
\end{eqnarray}
where
\begin{equation}\label{op1}
z=x+\ri y,\quad \bar z=x-\ri y,\quad p_z=\frac{1}{2}(p_x-\ri p_y),\quad p_{\bar z }=\frac{1}{2}(p_x+\ri p_y).
\end{equation}
We have set $\omega={eB}/{2c}$.
The operators (\ref{op1}) obey the basic commutation relations,
\begin{equation}
[x,p_x]=\ri\hbar 1\!\!1=[ y,p_{ y}],\quad [x,p_{ y}]=0=[y,p_{x}],
\end{equation}
\begin{equation}
[z,p_z]=\ri\hbar1\!\!1=[\bar z,p_{\bar z}],\quad [z,p_{\bar z}]=0=[\bar z,p_{z}].
\end{equation}
We use the formalism based on complex quantities to reduce the two-dimensional system to a one-dimensional system.
The creation and annihilation operators in the complex formalism are given as follows
\begin{equation}\label{r1}
 a=\ri\left(\frac{1}{\sqrt{\omega\hbar}}p_{\bar z}-\frac{\ri}{2}\sqrt{\frac{\omega}{\hbar}}z\right),\quad 
 a^\dagger=-\ri\left(\frac{1}{\sqrt{\omega\hbar}}p_{z}+\frac{\ri}{2}\sqrt{\frac{\omega}{\hbar}}\bar z\right).
\end{equation}
Consequently, the equation (\ref{eqq}) becomes
\begin{eqnarray}\label{r2}
 v\!\left(\begin{array}{cc}
		0& 2\ri\sqrt{\hbar\omega}a^\dagger\\
		 - 2\ri\sqrt{\hbar\omega}a& 0
	\end{array}\right)\left(\begin{array}{c}
		\varphi_1\\
		\varphi_2
		\end{array}\right)=E\left(\begin{array}{c}
		\varphi_1\\
		\varphi_2
		\end{array}\right).
\end{eqnarray}
Thus, the problem is transformed into a one-dimensional case whose eigenvalue is given by
\begin{equation}
 E_n=2v\sqrt{\hbar\omega n}.
\end{equation}

\section{Lorentz-covariant deformed algebra with minimal
length}

We review the Lorentz covariance problem in deformed algebra.
In the spacetime, coordinates are denoted by contravariant $(D+1)$-vectors $x^\mu=(x^0,x^i)$. The corresponding covariant $(D+1)$-vectors are given by $x_\mu=(x_0,x_i)$, with $g_{\mu\nu}=g^{\mu\nu}=\mbox{diag}(1,-1,-1,-1)$. Contravariant and covariant momenta are similarly defined as $p^\mu$ and $p_\mu$. The commutation relation of these operators are given by $[x_\mu, p_\nu]=-\ri\hbar g_{\mu\nu}$. The Lorentz covariance in the deformed algebra is given by
\begin{equation}\label{l}
 [X_\mu,P_\nu]=-\ri\hbar(1-\bar \beta P_\rho P^\rho)g_{\mu\nu},\quad [X_\mu,X_\nu]=2\ri\hbar\bar\beta(P_\mu X_\nu-P_\nu X_\mu), \quad [P_\mu,P_\nu]=0,
\end{equation}
where we assume that $\bar\beta={\beta}/{\hbar \omega}$ is a very small non-negative parameter. $\bar \beta$ has the dimension of an inverse squared momentum and $\beta$ is dimensionless.

The study of a system constructed from the  algebra (\ref{l}) is called the Synder model. This model is based on an algebra that includes spacetime coordinates and Lorentz generators. This is an example of noncommutative spacetime, meaning that the position operators have nontrivial commutation relations.

We can convert the covariant Lorentz algebra (\ref{l}) into a non-covariant Lorentz algebra,
\begin{equation}\label{crl}
 [X_i,P_j]=-\ri\hbar(1+\bar \beta P^2)g_{ij}, \quad [X_i,X_j]=2\ri\hbar\bar\beta(P_i X_j-P_j X_i), \quad [P_i,P_j]=0.
\end{equation}
In the momentum representation, the deformed position $X_i$ and momentum operators $P_i$ are represented by
\begin{equation}\label{op}
 X_i=\ri\hbar(1+\bar\beta p^2)\frac{\partial}{\partial p_i}+\ri\hbar \gamma p_i,\quad P_i=p_i,\quad [x_i, p_j]=-\ri\hbar g_{ij},
\end{equation}
where $\gamma$ is an arbitrary real constant, which does not effect the commutation relations (\ref{crl}). This constant, that affects the weighting function in the scalar product in the momentum space, ensures the Hermiticity of the operators (\ref{op}),
\begin{equation}
 \langle \psi|\phi \rangle=\int \frac{\rd^Dp}{(1+\bar\beta p^2)^{1-({\gamma}/{\bar\beta})}}\psi^\star(p)\phi(p).
\end{equation}
The algebra (\ref{l}) gives rise to nonzero minimal uncertainties in the position coordinates, hence
\begin{equation}
 (\Delta X) _i=(\Delta X)_0=\hbar\sqrt{D\bar\beta}.
\end{equation}

In the next section, we study the two-dimensional Dirac--Weyl equation in the presence of a minimal length in the case where the deformed algebra (\ref{l}) remains invariant under standard Lorentz transformations~\cite{C1}.

\section{Two-dimensional Dirac--Weyl equation in the  presence of a minimal length}

In this part, taking into account the results obtained in (\ref{r1}) and (\ref{r2}), we construct a natural extension with the same property, exhibiting the familiar structure of Fock algebra operators. Indeed, in the presence of a minimal length, the Fock algebra becomes
 $[{\cal A},{\cal A}^\dagger]=(1+\bar\beta p^2)$,
where the ladder operators ${\cal A}$ and~${\cal A}^\dagger$ in the momentum space are given by~\cite{dos2}, $\bar\omega=\sqrt{\hbar \omega}$,
\begin{eqnarray}\label{la}
 {\cal A}=\ri\frac{\bar\omega}{\sqrt 2}\left[\frac{p}{\bar\omega^2}+\big(1+\bar\beta p^2\big)\frac{\partial}{\partial p}\right], \quad
 {\cal A}^\dagger=\ri\frac{\bar\omega}{\sqrt 2}\left[-\frac{p}{\bar\omega^2}+\big(1+\bar\beta p^2\big)\frac{\partial}{\partial p}\right].
\end{eqnarray}
In terms of the ladder operators (\ref{la}), the Hamiltonian of the Dirac--Weyl in presence of the minimal length is give by
\begin{eqnarray}\label{hg}
	{\cal H}_{\beta}=\left(\begin{array}{cc}
		0& \lambda^\star {\cal A}^{\dagger}\\
		\lambda {\cal A}  & 0
	\end{array}\right),\quad \lambda=-2v\ri\sqrt{\hbar\omega}.
\end{eqnarray}
In the limit
$\bar\beta\longrightarrow 0$, we have
\begin{eqnarray}
 \lim_{\bar \beta\longrightarrow 0}{\cal A}=a=\ri\frac{\bar\omega}{\sqrt 2}\left(\frac{p}{\bar\omega^2}+\frac{\partial}{\partial p}\right),\quad
 \lim_{\bar \beta\longrightarrow 0}{\cal A}^\dagger=a^\dagger=
 \ri\frac{\bar\omega}{\sqrt 2}\left(-\frac{p}{\bar\omega^2}+\frac{\partial}{\partial p}\right).
\end{eqnarray}
We can obtain an interesting result if we define the operators ${\cal A}$ and
${\cal A}^\dagger$ as
\begin{eqnarray}\label{r1}
{\cal A}=\sqrt{\frac{\beta}{2}(a^\dagger a+2\ell)}a,\quad
{\cal A}^\dagger=a^\dagger\sqrt{\frac{\beta}{2}(a^\dagger a+2\ell)},
\end{eqnarray}
where, ${\cal A}\equiv {\cal J}_-$ and ${\cal A}^\dagger\equiv{\cal J}_+$
are the generators of the $su(1,1)$ algebra and satisfy the following
$su(1,1)$ algebra,
\begin{equation}
 [{\cal J}_-,{\cal J}_+]=2\sqrt{\frac{\beta}{2}}{\cal J}_0,\quad
 [{\cal J}_0,{\cal J}_\pm]=\pm\sqrt{\frac{\beta}{2}}
 {\cal J}_\pm,
\end{equation}
with
\begin{equation}\label{r2}
 {\cal J}_0=\sqrt{\frac{\beta}{2}}(a^\dagger a+\ell),\quad \ell=1+\sqrt{1+
 \frac{1}{\beta^2}}.
\end{equation}
We show that the realizations
(\ref{r1}) and (\ref{r2})
 particularly play a dominant role in the solution of the Hamiltonian (\ref{hg}). Indeed,
the action of the realizations (\ref{r1}) and (\ref{r2}) on the state
$|\ell,n\rangle$  ($n = 0, 1, 2,\dots$) leads
to infinite dimensional unitary irreducible representation of $su(1, 1)$ known as the positive representation~$D^+(\ell)$,
 \begin{align}
 {\cal J}_0|\ell,n\rangle&=\sqrt{\frac{\beta}{2}}{(n+\ell)}|\ell,n\rangle,\nonumber\\ 
 \;\;{\cal C}|\ell,n\rangle&=\frac{\beta}{2}\ell(\ell-1)|\ell,n\rangle,\\
 %\end{align}
%\begin{align}
 {\cal J}_-|\ell,n\rangle&=\sqrt{\frac{\beta}{2}n(2\ell+n-1)}|\ell,n-1\rangle,\nonumber\\
 {\cal J}_+|\ell,n\rangle&=\sqrt{\frac{\beta}{2}(n+1)(2\ell+n)}|\ell,n+1\rangle.
 \label{rec}
\end{align}
The eigenvalue problem of the Hamiltonian (\ref{hg}) gives
\begin{equation}
 {\cal H}_\beta\Psi=E^\beta\Psi,
\end{equation}
where $\Psi=(\psi_1,\;\psi_2)^T$. We get a set of coupled equations as follows
\begin{equation}\label{e1}
 E^\beta|\psi_1\rangle=\lambda{\cal J}_+|\psi_2\rangle,\quad
 E^\beta|\psi_2\rangle=\lambda^\star{\cal J}_-|\psi_1\rangle.
\end{equation}
Using the above equation, we have
\begin{equation}\label{e2}
 |\psi_2\rangle=\frac{{\lambda^\star}
 {\cal J}_-|\psi_1\rangle}{E^\beta}.
\end{equation}
By inserting equation (\ref{e2}) into (\ref{e1}), we get

\begin{equation}\label{ega}
 (E^\beta)^2|\psi_1\rangle=
 \lambda\lambda^\star{\cal J}_+{\cal J}_-|\psi_1\rangle.
\end{equation}
When we write the component $|\psi_1\rangle$ in terms of states
\begin{equation}
 |\ell,n\rangle=\sqrt{\frac{2^n\Gamma(2\ell)}{(\beta)^nn!\Gamma(n+2\ell)}}({\cal J}_+)^n|\ell,0\rangle,
\end{equation}
the equation (\ref{ega}) can be diagonalized and the energy spectrum is given by
\begin{equation}\label{ep}
 {E_n^\beta}^\pm=\pm\sqrt{\lambda\lambda^\star\frac{\beta}{2}n(2\ell+n-1)}.
\end{equation}
The total associated eigenstate is
\begin{eqnarray}\label{leq3}
	\Psi =\left(\begin{array}{c}
		1\\
		\displaystyle{\lambda^\star {\cal J}_-}/{E}
	\end{array}\right)|\ell,n\rangle.
\end{eqnarray}
It is clear that the energy of the system depends explicitly on the noncommutative parameter associated with the positions, and the zero-energy level for this system is $E^{\beta\pm}_0 = 0$. In the limit $\beta=0$, we obtain
\begin{equation}
 E^{0\pm}_n=2v\sqrt{\hbar\omega n},
\end{equation}
which exactly coincides  with the results found in \cite{aa1} and \cite{aa} for the special cases.

Our results are in agreement with those found in the literature. We found that the energy levels exhibit a dependence on $n^2$ in the presence of the minimum length, which describes a hard confinement. For example, in the limit $\beta= 0$, we find the results of the article \cite{Ba}.
In the work \cite{Ba1}, by passing into the commutative space, we find the same results for the case $\beta=0$. Moreover, our algebraic method generalizes the results found in \cite{Ba2}.

 The energy states are divided into two sectors: the positive sector made up of states of particles of positive energy and the negative sector made up of states of particles of negative energy. Additionally, negative energy levels are identified at antiparticle levels.  We can then assume that only positive energy particles are more appropriate to determine the thermodynamic properties of the system.

 \section{Thermal properties of the Dirac oscillator}
In order to study the thermodynamic properties of the Dirac--Weyl system, we start by defining the partition function of the system. Thus, given the positive energy spectrum~(\ref{ep}), the partition function~$Z$ at finite temperature $T$ is defined by
\begin{eqnarray} \label{leg}
	Z = \sum_{n=0} \re^{-\alpha E^\beta_n} , \quad \mbox{with} \quad \alpha = \frac{1}{k_{\rm B} T},
\end{eqnarray}
 where $k_{\rm B}$ is the Boltzmann constant, $E_n$ is the positive energy given by
 \begin{eqnarray}\label{en}
 {E_n^\beta}=\sqrt{\lambda\lambda^\star\frac{\beta}{2}n(2\ell+n-1)},\quad \lambda=-2v\ri\sqrt{\hbar\omega},\quad \omega=\frac{eB}{2c}.
\end{eqnarray}
Therefore, we rewrite the partition function (\ref{leg}) as follows
\begin{equation} \label{legg}
	Z = 1+\displaystyle\sum_{n=0}\exp\left({-\displaystyle\frac{1}{\rho}\displaystyle\sqrt{an^2+bn+1}}\right),
\end{equation}
where
\begin{equation}
 a=\frac{1}{2\ell},\quad b=1+a,\quad \rho=k_{\rm B}T\sqrt{\frac{c}{2\hbar eBv^2\beta \ell}},\quad \ell=1+\sqrt{1+\frac{1}{\beta}}.
\end{equation}
Using the formula \cite{y}
\begin{equation}\label{leq29}
	\re^{-x} = \frac{1}{2\piup \ri }\int_{C} \rd s\, x^{-s}\Gamma(s),
\end{equation}
the partition function(\ref{legg}) can take the following form dependent on the Epstein zeta function,
  \begin{equation}
   Z=1+I_1+I_2+I_3,
  \end{equation}
where
\begin{equation}\label{zpp}
	\displaystyle I_1=\displaystyle\frac{1}{2\piup \ri}\int_{C} \rd s \left(\frac{1}{\rho}\right)^{-s} 2 (a)^{-\frac{s}{2}}\zeta(s)\Gamma(s),
\end{equation}
\begin{equation}
	I_2\!=\!\displaystyle\frac{1}{2\piup \ri} \int_{C} \rd s \left(\frac{1}{\rho}\right)^{-s} 2 (a)^{-\frac{s}{2}}
		y^{\left(\frac{1}{2} - \frac{s}{2}\right)} \frac{\sqrt{\piup}}{\Gamma\left(\frac{s}{2}\right)} \zeta (s-1) \Gamma \left(\frac{s}{2} - \frac{1}{2}\right) \Gamma (s),
    \end{equation}	
\begin{equation}	
	I_3=\displaystyle\frac{1}{2\piup \ri}\int_{C} \rd s\left(\frac{1}{\rho}\right)^{-s} 2 \frac{a^{-\frac{s}{2}}\,
        y^{\frac{s}{2} - \frac{1}{2}} }{\Gamma \left(\frac{s}{2}\right)} \piup^{\frac{s}{2}}\, H{\left(\frac{s}{2}\right)}
    \Gamma (s).
\end{equation}
$\Gamma (s)$ is the Euler function, $\zeta (s)$ is zeta
function. $H\left(\frac{s}{2}\right)$
is the power function given by
\begin{equation}
	H\left(\frac{s}{2}\right)= 4 \sum_{k=1}^{N} \sigma_{(1-s)} k^{\frac{s}{2} - \frac{1}{2}} \cos(2\piup kx)\,
	K_{\frac{s}{2} - \frac{1}{2}}\,2k\piup y,\nonumber\\
\end{equation}
where $K_{\frac{s}{2} - \frac{1}{2}}$ is the Bessel function and
\begin{equation}\label{con}
 \sigma_{_{(1-s)}} = \sum_{k/m} \left(\frac{k}{m}\right)^{1 - s}, \quad
x=\frac{b}{2a},\quad y =  \frac{\sqrt{b^2-4a}}{2a},\quad b^2-4a>0.
\end{equation}
Thus,
the integral $I_1$ has two poles in $s = 0$ and $s = 1$. The integral $I_2$ has three poles in $s = 0,\; s = 1$
and $s = 2$, and the integral
$I_3$ has a pole at $s = 0$. By applying the residues theorem and using the following limit,
\begin{equation}
 \displaystyle\lim_{s\rightarrow 0}[s \zeta(s) \Gamma(s)]=-\frac{1}{2},\quad 
 \lim_{s\rightarrow1} \left[(s-1) \zeta(s)\Gamma(s)\right]=1.
 \end{equation}

\begin{equation}
  \lim_{s\rightarrow1}\Bigg[\frac{(s-1)\zeta(s-1)\Gamma\big(\frac{s}{2}-\frac{1}{2}\big)\Gamma(s)}{\Gamma\left(\frac{s}{2}\right)}\Bigg]=-\frac{1}{\sqrt{\piup}},
\end{equation}

\begin{equation}
 \displaystyle\lim_{s\rightarrow0} \left[\frac{s H\left(\frac{s}{2}\right)\Gamma(s)}{\Gamma\left(\frac{s}{2}\right)}\right]=0,
 \end{equation}
we get
\begin{equation}
I_1=-1+2\frac{\rho}{\sqrt{a}},\quad I_2=-2\frac{\rho}{\sqrt{a}}+
 \frac{2\piup \rho^2}{a y},\quad  I_3=0.
\end{equation}
Consequently, we obtain the partion function in the following form,
\begin{equation}\label{pf}
 Z=\frac{2\piup \rho^2}{a y},\quad y=\frac{1}{2}(2\ell-1).
\end{equation}
From the equation (\ref{en}), we can rewrite the above equation
\begin{equation}\label{pf}
 Z=\frac{8\piup}{\beta(2\ell-1)}\frac{1}{\tau},
\end{equation}
with
\begin{equation}
 \tau=\alpha \delta,\quad \alpha=\frac{1}{k_{\rm B}T},\quad \delta=v\sqrt{\frac{2\hbar eB}{c}}.
\end{equation}
Although in principle the deformation parameter $\beta$ is not fixed by theory, it is generally assumed to be of the order of unity. This is particularly the case in some models of string theory \cite{r1a,r1b}. Many studies have appeared in the literature, with the aim of setting experimental limits to $\beta$ (see \cite{r2,r3} and the references therein). Different theoretical frameworks, and an explicit analytical calculation~\cite{r4}, indicate a value of $\beta$ of order $1$, in particular $\displaystyle\beta\simeq{82\piup}/{5}$.
In this work the value of $\beta$ has not been fixed.  It can take on all positive values, i.e., $\beta \geqslant 0$.
This is contrary to the results found for our previous work on the
deformed Dirac model~\cite{dos2}. The partition function well delimits the values of the deformation parameter.

Once the domain of definition of our deformation parameter $\beta$ is well specified, we can determine from the partition function (\ref{pf}) all the thermodynamic quantities such as the total energy, the free energy, the entropy and the specific heat:

\begin{itemize}[label={--}]
	\item {the total energy}
	\begin{eqnarray}
		\frac{U}{\delta} = -\frac{\partial}{\partial \tau} \ln( Z) =  \frac{2}{\tau};
	\end{eqnarray}
	\item {the free energy}
	\begin{eqnarray}
		%\begin{array}
		\frac{F}{\delta}&=&-\frac{1}{\tau}\ln (Z) = -\frac{1}{\tau} \ln \left[\frac{8\piup}{\beta(2\ell-1)}\frac{1}{\tau^2}\right];
		%\end{array}
	\end{eqnarray}
	
	\item {the entropy}
	\begin{eqnarray}
		\frac{S}{k_{\rm B}} = \tau^2 \frac{\partial}{\partial \tau }\left(\frac{F}{\delta}\right)
		=  \ln \left[\frac{8\piup}{\beta(2\ell-1)}\frac{1}{\tau^2}\right]+2;
	\end{eqnarray}
	
	\item {the specific heat}
	\begin{eqnarray}
		\frac{C}{k_{\rm B}} = -\tau^2 \frac{\partial}{\partial \tau}\left(\frac{U}{\delta}\right)=  2.
	\end{eqnarray}
	
\end{itemize}

We note that the average energy and specific heat are independent of the deformation parameter. Moreover, the Dulong--Petit law for an ultra-relativistic ideal gas is verified. It is worth mentioning that similar results were observed for the Dirac and Kemmer oscillators in a thermal bath \cite{15a,16a}.
It was also found in \cite{aa1a} that in the high temperature region the dominant term of the Dunkl partition function becomes independent of the Wigner parameter.

\begin{figure}[htbp]
	\begin{center}
		\includegraphics[scale=0.65]{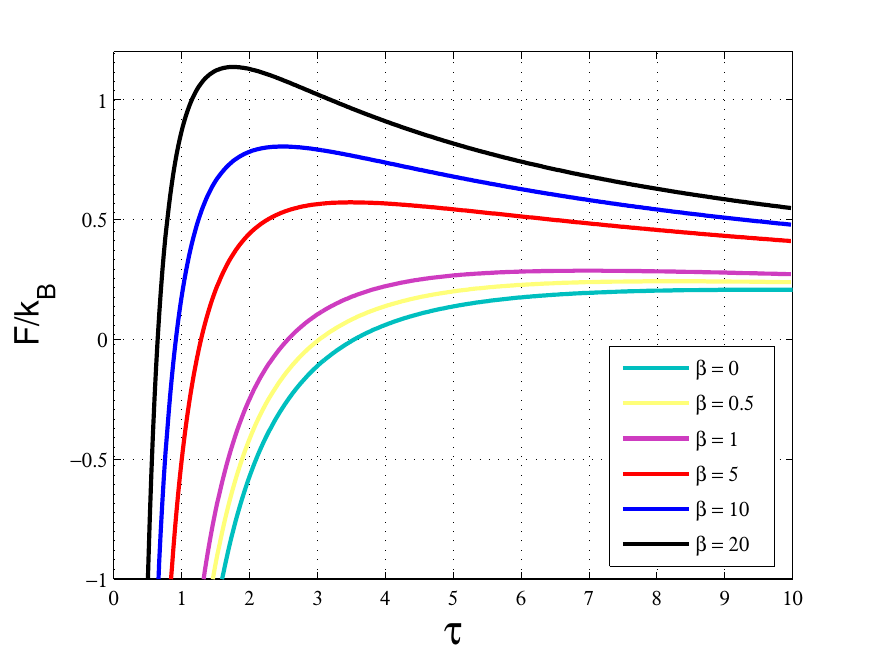}
	\end{center}
	\caption{(Colour online) Free energy as a function of temperature for different values of parameter $\beta$.}
	\label{wide1}
\end{figure}

The analysis of these figures shows us that the deformation parameter $\beta$ plays an important role on the thermodynamic properties. Indeed,  figure~\ref{wide1} shows the free energy as a function of temperature for different values of the parameter $\beta$. It is clear from the figure that the free energy increases until it reaches a maximum, then decreases with a decreasing temperature and shifts to positive values. We notice that at each curve there is a peak that appears over a small temperature range. This peak is more noticeable when the deformation parameter is higher and then all these curves go towards an increasing temperature of constant value.

\begin{figure}[htbp]
	\begin{center}
		\includegraphics[scale=0.65]{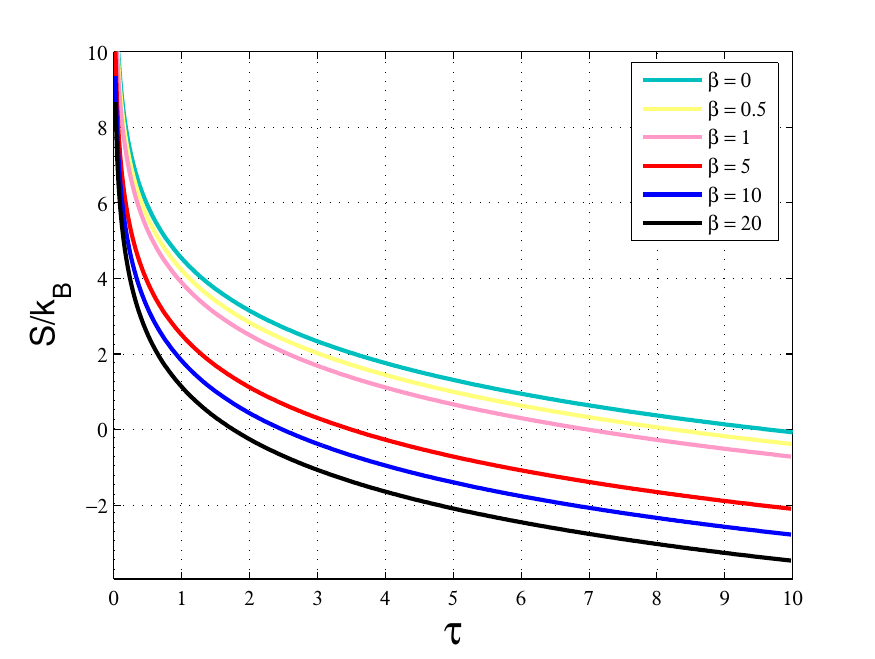}
	\end{center}
	\caption{(Colour online) Entropy as a function of temperature for different values of parameter $\beta$.}
	\label{wide2}
\end{figure}

Figure~\ref{wide2}  shows entropy as a function of temperature for different parameters $\beta$. As expected, entropy decreases with a decreasing temperature at a fixed $\beta$ parameter. At fixed values of temperature, the entropy decreases with an increasing parameter $\beta$. Thus, the reduction of the system is an order.

Since the specific heat tends to an asymptotic value equal to $2 k_{\rm B}$, we can then say that the specific heat of two-dimensional graphene coincides with that of the two-dimensional harmonic oscillator for higher temperatures. Moreover, as in the non-relativistic case \cite{cc}, we can argue this situation by saying that these limits follow the Dulong--Petit law for a relativistic ideal gas.

We have solved the Dirac-Well equation algebraically within the framework of the generalized Heisenberg principle. The results obtained compare well with those obtained in the literature for particular cases~\cite{lm}. It has been recently shown that the graphene system can be modulated under a uniform magnetic field. We have determined the thermodynamic properties of the deformed graphene in the presence of a minimum length using the Epstein zeta function. We have found that the algebraic solution does not impose any condition on the deformation parameter. This solution method confirms only the positive value of the deformation parameter.

\section{Conclusion}
We have studied the thermodynamic properties of a massless Dirac electron in graphene subjected to a uniform magnetic field with the generalized Heisenberg uncertainty principle. This principle, through the deformed Heisenberg algebra, brought out the hidden symmetries of the $su(1,1)$ Lie algebra of the system. The eigenvalues and eigenstates have been constructed algebraically and form the infinite-dimensional representation of the deformed $su(1, 1)$ algebra.
These are used together with a method based on the zeta function to explicitly determine the partition function. Thus, thermodynamic functions such as Helmholtz free energy, total energy, entropy and heat capacity have been obtained in terms of the deformation parameter $\beta$. The overall behavior of the thermodynamic functions indicates that the deformation parameter produces a deviation around the usual undeformed profile. We note that this deviation for a fixed temperature is positive for the free energy and negative for the entropy. It is both interesting and surprising to find that the specific heat and the average energy are independent of the deformation parameter.
Finally, we hope that in the near future, as soon as high-precision experimental measurements involving the thermodynamic properties of graphene are obtained, our results can be used as a good tool to study these properties. Furthermore, we intend to investigate the possibility that new features of graphene can be described by more general models with the higher-order generalized uncertainty principle.

\newpage

\ukrainianpart

\title{Алгебраїчний розв'язок та термодинамічні властивості графену за наявності мінімальної довжини}
\author{Дж. Ґеботого\refaddr{label1}, Ф. А. Досса\refaddr{label1},
	Г. Й. Х. Авоссеву\refaddr{label2}}
\addresses{
	\addr{label1} Лабораторія прикладної фізики, Національний університет наук, технологій, інженерії та математики (UNSTIM) Абомей, BP: 2282 Гохо Абомей, Бенін
	\addr{label2} Інститут математики та фізичних наук,
	Університет Абомей-Калаві,
	01 BP 613 Порто-Ново, Бенін}
%
%% якщо автор є один або автори є з однієї установи:
%
%  \author{1й Автор, 2й Автор, \ldots}
%  \address{Інститут\ldots}
%
%%

\makeukrtitle

\begin{abstract}
	\tolerance=3000%
	Графен --- це напівпровідник з нульовою забороненою зоною, де електрони, що поширюються всередині, описуються ультрарелятивістським рівнянням Дірака, яке зазвичай використовується для безмасових частинок дуже високої енергії. У цій роботі ми показуємо, що графен під дією магнітного поля за наявності мінімальної довжини має приховану симетрію $su(1,1)$. Ця симетрія дозволяє нам побудувати спектр алгебраїчно. Фактично, узагальнене співвідношення невизначеності, що призводить до ненульової мінімальної невизначеності положення, було б ближчим до фізичної реальності та дозволило б нам контролювати або створювати зв'язані стани в графені. Використовуючи функцію розподілу на основі дзета-функції Епштейна, отримуються задовільні значення термодинамічних характеристик. Виявлено, що закон Дюлонга--Пті добре підтверджується, а теплоємність не залежить від параметра деформації.

	
\end{abstract}

\lastpage
\end{document}